\documentclass{egpubl}
\usepackage{egsgp2024}
 
\SpecialIssuePaper         

\CGFccby

\usepackage[T1]{fontenc}
\usepackage{dfadobe}  

\biberVersion
\BibtexOrBiblatex
\usepackage[backend=biber,bibstyle=EG,citestyle=alphabetic,backref=true]{biblatex} 
\electronicVersion
\PrintedOrElectronic

\ifpdf \usepackage[pdftex]{graphicx} \pdfcompresslevel=9
\else \usepackage[dvips]{graphicx} \fi

\usepackage{egweblnk} 
\newcommand{\refequ}[1]{Eq.~(\ref{equ:#1})}

\newcommand{\reffig}[1]{Figure~\ref{fig:#1}}
\newcommand{\reftab}[1]{Table~\ref{tab:#1}}
\newcommand{\refsec}[1]{Section~\ref{sec:#1}}

\providecommand{\A}{}
\providecommand{\B}{}
\providecommand{\C}{}
\providecommand{\D}{}

\providecommand{\F}{}
\providecommand{\G}{}

\providecommand{\I}{}
\providecommand{\J}{}

\providecommand{\L}{}
\providecommand{\M}{}
\providecommand{\N}{}

\providecommand{\R}{}
\providecommand{\S}{}

\providecommand{\Y}{}

\providecommand{\c}{}

\providecommand{\f}{}

\providecommand{\o}{}

\providecommand{\v}{}

\usepackage{tabularx}
\usepackage{booktabs}
\usepackage{makecell}
\usepackage[table]{xcolor}
\definecolor{white}{rgb}{1,1,1}
\definecolor{lightbluishgrey}{rgb}{0.76471,0.84824,0.91647}

\usepackage{subcaption}
\usepackage{wrapfig}
\usepackage{graphicx}

\usepackage{listings}
\usepackage{layouts}
\makeatletter 
\newcommand{\layoutdetails}{%
\begin{tabular}{ll}
 \texttt{\textbackslash{textwidth}} & \printinunitsof{in}\prntlen{\textwidth} \\
\texttt{\textbackslash{linewidth}} & \printinunitsof{in}\prntlen{\linewidth} \\
Main text font &  \f@size pt \f@family \\
\sffamily \small Caption text font &  \sffamily \small \f@size pt \f@family \\
\end{tabular}%
}
\makeatother 

\usepackage{amsfonts}
\usepackage{amsmath}
\usepackage{cancel}
\usepackage{extarrows}
\usepackage{centernot}
\usepackage{mathtools}
\usepackage{dsfont}
\usepackage{caption}
\renewcommand{\A}{\mathbf{A}}
\renewcommand{\J}{\mathbf{J}}
\renewcommand{\B}{\mathbf{B}}
\renewcommand{\C}{\mathbf{C}}
\renewcommand{\D}{\mathbf{D}}
\renewcommand{\F}{\mathbf{F}}
\renewcommand{\G}{\mathbf{G}}
\renewcommand{\I}{\mathbf{I}}
\renewcommand{\L}{\mathbf{L}}

\renewcommand{\M}{\mathbf{M}}

\renewcommand{\R}{\mathbb{R}}
\renewcommand{\S}{\mathbf{S}}
\renewcommand{\f}{\mathbf{f}}
\renewcommand{\c}{\mathbf{c}}
\renewcommand{\o}{\mathbf{o}}
\renewcommand{\v}{\mathbf{v}}
\newcommand{\dual}[1]{\star_{#1}}
\newcommand{\primal}[1]{\bullet_{#1}}

\DeclareSymbolFont{FooFont}{OT1}{cmr}{m}{n}
\DeclareMathSymbol{\fooEqual}{\mathrel}{FooFont}{`=}

\definecolor{darkgreen}{rgb}{0.13, 0.55, 0.13}

\newcommand{\newhl}[1]{#1}
\newcommand{\final}[1]{#1}

\renewcommand{\Y}{\includegraphics[trim=0 0.2em 0 -0.2em,height=1em]{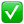}}
\renewcommand{\N}{\includegraphics[trim=0 0.2em 0 -0.2em,height=1em]{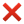}}

\newcounter{originalsecnumdepth}
\newcommand{\unnumberedsubsection}[1]{
  \setcounter{originalsecnumdepth}{\value{secnumdepth}}
  \setcounter{secnumdepth}{1}
  \subsection{#1}
  \setcounter{secnumdepth}{\value{originalsecnumdepth}}
}

\usepackage{layouts}

\title[Optimized Dual-Volumes for Tetrahedral Meshes]{Optimized Dual-Volumes for Tetrahedral Meshes}
\author[Jacobson]
{\parbox
    {\textwidth}
        {\centering 
        Alec Jacobson
        }
        \\
{\parbox
    {\textwidth}
        {\centering 
            University of Toronto and Adobe Research Toronto
        }
    }
}

\begin{document}
\maketitle
\begin{abstract}
  Constructing well-behaved Laplacian and mass matrices is essential for tetrahedral mesh processing.
  Unfortunately, the \emph{de facto} standard linear finite elements exhibit
  bias on tetrahedralized regular grids, motivating the development of finite-volume
  methods.
  In this paper, we place existing methods into a common construction, showing
  how their differences amount to the choice of simplex centers.
  These choices lead to satisfaction or breakdown of important properties:
  continuity with respect to vertex positions, positive semi-definiteness of the
  implied Dirichlet energy, positivity of the mass matrix, and unbiased-ness on regular grids.
  Based on this analysis, we propose a new method for constructing dual-volumes
  which explicitly satisfy all of these properties via convex optimization.
\end{abstract}

\begin{figure}
\includegraphics[width=\linewidth]{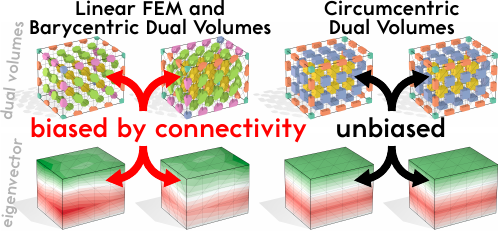}
  \caption{
    \label{fig:grid-3d}
    Linear FEM corresponds to dual-volumes formed by connecting simplex barycenters
  (top, volumes scaled by 50\%), but exhibit bias
  on tetrahedralized grids, e.g., a generalized eigenvector $\L \v_5 = 
  \lambda_5 \M \v_5$ with a pinned base (bottom).
  Our optimized dual-volumes coincide with circumcentric dual-volumes on a
  tetrahedralized regular grid (right), avoiding bias.
  }
\end{figure}

\begin{table}
  \centering
  \begin{tabular}{lcccccccccccccc}
                                     & $C^0$  & $-(\L+\L^T) \succeq 0$  & $\ \M > 0\ $ & \textsc{Grid}\\
    \hline                                                                          
    Barycentric                      & \Y     & \Y                   & \Y       & \N           \\
    Circumcentric                    & \Y     & \N                   & \N       & \Y           \\
    \hspace{1sp}\cite{AlexaHKS20}    & \N     & \N                   & \Y       & \N           \\
    \hspace{1sp}\cite{mullen2011hot} & \N     & \N                   & \N       & \Y           \\
    \hline                                                                
    Ours, snapping                   & \Y     & \N                   & \Y       & \Y           \\
    \textbf{Ours, optimized}         & \Y     & \Y                   & \Y       & \Y           \\
    \hline
  \end{tabular}
  \caption{
    \label{tab:feature}
    \newhl{
    Given a mesh with prescribed positions and connectivity we consider
    existing finite-volume methods.
    Are their resulting dual volumes continuous functions of vertex positions ($C^0$)?
    Do they 
    yield a positive semi-definite Dirichlet energy ($-(\L+\L^T) \succeq 0$)
    and generate a positive mass matrix ($\M > 0$)?
    Do they maintain unbiased performance on a
    tetrahedralized regular grid (\textsc{Grid})?
  }
  }
\end{table}

\section{Introduction}
While the finite-element method with piecewise-linear shape functions may be the
\emph{de facto} standard for tetrahedral mesh processing of solids,
finite-volume methods are useful for a variety of reasons.
In particular, unlike linear finite elements they can result in
unbiased operators for tetrahedralized regular grids (see \reffig{grid-3d}).
At the heart of the finite-volume method is the definition of local dual-volumes around each vertex of a given tetrahedral mesh.
%
Existing methods attempt to generalize orthogonal Voronoi dual-volumes, which behave well for Delaunay meshes.
Unfortunately, when considering general (e.g., non-Delaunay) meshes, each
approach gives up one or more desirable properties of its resulting Laplacian
and mass matrices: 
continuity with respect to vertex positions, 
positive semi-definiteness of its implied Dirichlet energy,
positivity of the mass matrix, unbiasedness
on regular grids (see \reftab{feature}). 

In this paper, we categorize existing methods according to these criteria.
Placing methods into a common construction, their differences amount to the
choice of simplex centers, most importantly triangle centers.
Each choice drastically affects the discretization of the divergence operator in a finite-volume construction.
From their constructions we show how properties of each method arise or break down algebraically.

\begin{figure}[t!]
\includegraphics[width=\linewidth]{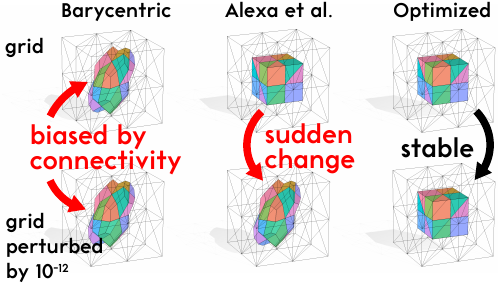}
  \caption{
    \label{fig:grid-continuity}
    Previous methods such as \cite{AlexaHKS20} reproduce circumcentric duals for
    perfect grids, where tetrahedra are on the cusp of becoming non-acute.
    A gentle nudge and their dual volumes jump to totally different shapes and sizes.
    Our optimized dual volumes are continuous functions of mesh vertex positions.
  }
\end{figure}

Based on this analysis, we propose a new method for constructing dual-volumes
which explicitly satisfies all of these properties via convex optimization.
In contrast to previous methods, our dual-volumes are continuous functions of
vertex positions (see \reffig{grid-continuity}).
As a warm-up, we consider modifying the positive-volume filtering
procedure of Alexa et al.~\cite{AlexaHKS20} to ensure continuity as tetrahedra morph from
acute to non-acute shapes (``Ours snapping'' in \reftab{feature}).
Unfortunately, this --- like \cite{AlexaHKS20} --- does not ensure
that the Laplacian matrix defines a positive semi-definite Dirichlet energy.
%
Instead, we propose directly optimizing the centers associated with each simplex
to explicitly ensure symmetry and negative semi-definiteness of the Laplacian.
All properties are satisfied by construction, including continuity as the
optimization problem is strictly convex.

Let's plunge into our problem's details. We will consider prevalent
prior methods after introducing the details of our desired properties and
postpone a broader discussion of related work until our final discussion in \refsec{discussion}.

%

\section{Finite Volume Operators}
%
%
Given a tetrahedral mesh of a solid region $\Omega \subset \R^3$ with $n$ vertices, 
we will follow a general finite-volume construction to build matrices $\M \in
\mathbb{R}^{n \times n}$ and $\L \in \mathbb{R}^{n \times n}$.

\subsection{Mass-matrix}
Our goal is to compute a diagonal or \emph{lumped} mass matrix $\M \in
\mathbb{R}^{n \times n}$, which as a quadratic form computes the inner product
of functions defined by per-vertex scalar values $\f \in \R^n$.
\begin{align}
  \f^T \M \f \approx \int_{\Omega} f^2 \, dV
\end{align}
We can also consider $\M$ as a linear operator acting on a \emph{yet to be
determined} local dual-volume $\dual{i} \subseteq \Omega$ around each vertex $i$:
\begin{align}
  (\M \f)_i \approx \int_{\dual{i}}  f \, dV 
\end{align}

To ensure the diagonally of $\M$ we now assume that $f$ is constant over each local volume $\dual{i}$.
This local dual volume $\dual{i}$ may further be broken into contributions within
the primal volume $\primal{t}$ of each tetrahedron $t$ (usually only those incident on $i$):

\begin{align}
  (\M \f)_i = \left( \sum_{t} \int_{\primal{t} \cap \dual{i}}  \, dV \right) \f_i,
  \quad
  \M_{ii} := \sum_{t} \int_{\primal{t} \cap \dual{i}}  \, dV,
\end{align}
with $\M_{ij} = 0$ for $i \neq j$.

\subsection{Dirichlet Energy / Laplacian Matrix}
Our goal is also to compute a sparse Laplacian matrix $\L \in \mathbb{R}^{n \times
n}$ which approximates the continuous Laplacian operator acting on scalar
functions or equivalently acting as a quadratic form computing the integral of
squared gradients:
\begin{align}
  -\f^T \L \f \approx \int_{\Omega} \nabla f \cdot \nabla f \, dV = -\int_{\Omega} f \Delta f \, dV
  \cancelto{0}{+ \int_{\partial \Omega} f \nabla f \cdot \hat{n} \, dS}
\end{align}
where the boundary term vanishes because we assume zero Neumann boundary
conditions ($\nabla f \cdot \hat{n} = 0$), which are this energy's natural boundary conditions.

Again, we consider the matrix as computing the local integral at each vertex
$i$:
\begin{align}
  (\L \f)_i \approx \int_{\dual{i}} \Delta f \, dV
\end{align}

This time, we assume that gradient $\nabla f$ is constant over each local dual volume $\dual{i}$, which will result in a sparse (but not diagonal) matrix $\L$.
Again the local dual volume $\dual{i}$ may be broken into contributions within
each primal tetrahedron volume $\primal{t}$, upon which divergence theorem may be applied:
\begin{align}
  \label{equ:continuous-div}
  (\L \f)_i = \sum_{t} \int_{\primal{t} \cap \dual{i}} \Delta f \, dV = 
  \sum_{t} \left(\int_{\partial (\primal{t} \cap \dual{i}) \cap\, \Omega} \hat{n} \, dS \right) \cdot \left(\nabla f\right)_t, 
\end{align}
where the ``$\cap\,\Omega$'' enforces the zero Neumann boundary conditions by \emph{not} including any part of the boundary of the overall solid volume $\Omega$.
By now $\L_{ij}$ are fully determined by the choice of local dual volume
$\dual{i}$, but we momentarily postpone its explicit formula until after
defining our general dual volume construction.

\section{Dual Volumes}

\begin{figure}
  \includegraphics[width=\linewidth]{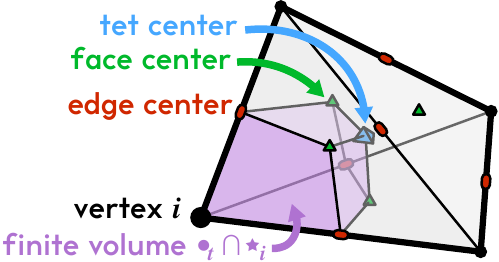}
  \caption{
  \label{fig:tet-centers}
    In our general construction, each tetraheron $t$ contributes a portion of its volume
    $\primal{t} \cap \dual{i}$ to the local finite dual-volume of each incident vertex $i$. This subvolume is determined by centers
    defined for each simplex. Different methods choose different centers leading to different properties of constructed operators.
    }
\end{figure}

In this paper, we provide a general construction of the local volume
contribution $\primal{t} \cap \dual{i}$ from each tetrahedron $t$ to the dual
volume of each mesh vertex $i$ as a hexahedron with vertices at ``centers'' of all
shared simplices (see
\reffig{tet-centers}):
\begin{itemize}
  \setlength{\itemindent}{1em}
  \item[] 1 at the vertex $i$,
  \item[] 3 at edges of $t$ incident on $i$,
  \item[] 3 at triangle facets of $t$ incident on $i$, and
  \item[] 1 at the tetrahedron $t$.
\end{itemize}

For each tetrahedron $t$ incident on a vertex $i$, these centers define the
eight corners of a hexahedron $\primal{t} \cap \dual{i}$.
To compute the entries of $\M$ and $\L$, we may explicitly create a mesh of this local contribution and compute its volume (for $\M$) or integral of surface normals (for $\L$).
For simple choices of centers, we can cook up closed-form expressions for the entries of $\M$ and $\L$ without explicitly realizing a mesh of $\primal{t} \cap \dual{i}$.

Nevertheless, the choice of how each center is defined will affect the following criteria we
may care about in the mass and Laplacian matrix definitions above:

\unnumberedsubsection{$C^0$ --- continuous function of vertex positions}
The local volume $\dual{i}$ should be a continuous function of tetrahedral mesh vertex positions.
This ensures that integrated quantities stored on mesh vertices don't jump in
value as the underlying mesh changes continuously, for example, in Lagrangian
simulation or mesh deformation methods (see \reffig{grid-continuity}).

\begin{figure}[t!]
\includegraphics[width=\linewidth]{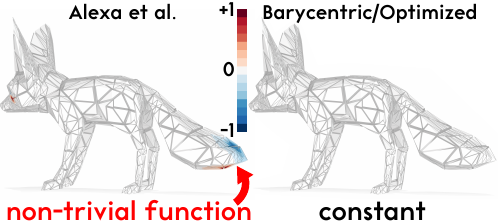}
  \caption{
    \label{fig:fox}
    In this numerical example, there exists a non-trivial unit-norm function $\hat{\f}$ (left) such that
    the implied Dirichlet energy of \cite{AlexaHKS20} is negative (and the energy is
    nonconvex): $\hat{\f}^T (-\L) \hat{\f} \approx -2.96$.
    For the barycentric and our optimized centers, $\L$ is symmetric negative
    semi-definite and that the smallest energy value is zero from a
    constant function (right).
  }
\end{figure}

\unnumberedsubsection{$-(\L + \L^T) \succeq 0$ --- positive semi-definite Dirichlet energy}
%
\newhl{Treating the Laplacian matrix $\L$ as a quadratic form should result in a 
semi-definite\footnote{\newhl{We use $\mathbf{X} \succeq 0$ to write that the matrix $\mathbf{X}$ is positive semi-definite and $\mathbf{z} \ge 0$ to write that every element of $\mathbf{z}$ is non-negative.}} Dirichlet energy.}
That is, for any input scalar function $\f \in \R^n$ we expect a non-negative Dirichlet energy:
\begin{align}
  -\f^T \L \f \ge 0 \quad \forall\, \f \in \R^n,
\end{align}
with equality only for constant functions (see \reffig{fox}).
Various methods (e.g.,
\cite{ValletL08,Rustamov07,SunOG09,OvsjanikovBSBG12,SharpACO22,BenchekrounZCGZJ23})
rely on explicit spectral decomposition and Dirichelt energy indefiniteness presents a
major problem.

Dirichlet energy positive semi-definiteness requires that the sum of $\L$ and
its transpose is negative semi-definite or equivalently that all eigenvalues of
the Laplacian plus its transpose are real and non-positive:
\begin{align}
  \lambda_{\L+\L^T} \le 0,
\end{align}
with a single zero eigenvalue corresponding to the constant function.
While symmetry and negative semi-definiteness together are \emph{sufficient} to
ensure this property, they are not \emph{necessary}.
\begin{align}
  \L = \L^T \quad \text{and} \quad \lambda_{\L} \le 0 \quad 
  \substack{\Longrightarrow \\[0.5em] \centernot\Longleftarrow}
  \quad -( \L + \L^T) \succeq 0.
\end{align}
Indeed, symmetry alone or non-positive eigenvalues alone does not appear to 
ensure any practical value of $\L$.
Meanwhile, an indefinite $\L$ which does define a positive semi-definite
Dirichlet energy is effective for many or most applications.

The risk of losing Dirichlet energy positive semi-definiteness is an immediate
consequence of the choice to discretize at the operator level rather than the
quadratic form level.
Alternatively, linear FEM or discrete calculus of variations would alternatively
discretize the function space and operators present in the energy functional
itself, where all terms are squared and thus non-negative.

Positive semi-definite or not, we can consider the Dirichlet ``energy'' implied by $\L$ as a
summation of contributions per-tetrahedron:
\begin{align}
  \f^T \L \f = \sum_t \f^T\S_t^T \L_t \S_t \f,
\end{align}
where $\S_t \in \R^{4 \times n}$ selects the entries of $\f$ corresponding to
the vertices of tetrahedron $t$, and $\L_t \in \R^{4 \times 4}$ is the
contribution of tetrahedron $t$ to the Laplacian matrix.
By assuming that $\nabla f$ is constant in each tetrahedron $t$, we have reduced
the discretization of the integrated Laplacian to that of the per-vertex divergence
operator, which in turn can be decomposed into a sum of contributions per-tetrahedron, namely:
\begin{align}
  \L_t &= \D_t \underbrace{\G_t}_{\mathrlap{\text{fixed by assumption}}}
\end{align}
where $\D_t \in \R^{4 \times 3}$ and $\G_t \in \R^{3 \times 4}$ and the entries are determined according to 
\refequ{continuous-div}:
\begin{align}
  \label{equ:discrete-div}
  \left(\S_t^T \D_{t}\right)_i = \int_{\partial (t \cap \dual{i}) \cap\, \Omega} \hat{n} \, dS  \, 
\end{align}

%
If we require symmetry $\L = \L^T$ for all input meshes, we are --- by reduction --- requiring that:
\begin{align}
  \label{equ:symmetry-constraint}
  \D_t \G_t = \G_t^T \D_t^T = -\G_t \A \G_t^T 
\end{align}
for all tetrahedra $t$, where $\A_t \in \R^{3 \times 3}$ is some symmetric matrix
and minus sign appears to foreshadow the interpretation of $\A_t$ as a
piecewise-constant metric or diffusion tensor.
%
\newhl{Letting $\mathds{1}$ be the vector of ones, then 
because $\G_t \mathds{1}=0$, we immediately have $\L_t \mathds{1} = 0$}. If we
further require that $\L_t \preceq 0$, then we are requiring that $\A_t$ is positive definite:
\begin{align}
  \label{equ:A-definite-constraint}
  \A_t \succ 0.
\end{align}
%


\unnumberedsubsection{$\M > 0$ --- positive mass matrix}
The mass matrix $\M$ should be positive for a non-degenerate input mesh.
As $\M_{ii}/\sum_{j} \M_{jj}$
represents the percentage of the total volume of $\Omega$, we expect each local
volume $\dual{i}$ to be positive.
Since $\M$ is diagonal by construction, this property is true if and only if the
mass-matrix $\M$ is strictly positive definite.
Indefinite or worse singular $\M$ causes all sorts of problems in numerical methods.
A sufficient condition for this property is that the local volume contribution
in each tetrahedron is positive: $\int_{\primal{t} \cap \dual{i}} \, dV > 0$ for
$t \ni i$, or more strictly that all centers lie within their respective simplices.

\unnumberedsubsection{\textsc{Grid} --- unbiased volumes on regular grids}
As a unit-test for ``connectivity bias,'' the dual volume definition should
result in unbiased local volumes on a regular grid where each cube cell is
subdivided into six tetrahedra around an arbitrary diagonal of symmetry (see \reffig{grid-3d}).

With conditions in hand, let's immediately consider some obvious center choices and then those of prevalent prior works (summarized in \reftab{feature}).

\subsection{Barycentric dual-volumes}
The barycentric dual-volumes are defined by using the barycenter (a.k.a.,
centroid or center of mass) of each simplex.
For the mass-matrix, this results in an equal division of each tetrahedron volume to its incident vertices:
\begin{align}
  \M^\text{bary}_{ii} = \frac{1}{4} \sum_{t \ni i} \int_{\primal{t}} dV
\end{align}

Using barycenters results in the same Laplacian matrix as direct application of
the linear finite element method: the ubiquitous cotangent Laplacian.

\begin{figure}[b!]
\includegraphics[width=\linewidth]{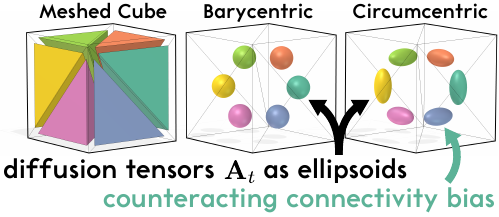}
  \caption{
    \label{fig:ellipsoids}
    In our construction, any choice of dual-volume implies a $3\times 3$ matrix
    $\A_t$ which can be interpreted as a piecewise-constant diffusion tensor.
    Barycentric dual-volumes, corresponding to linear FEM, imply tensors
    proportional to the identity: in a way, \emph{trusting} the connectivity too much.
    Circumcentric dual-volumes imply tensors which counteract connectivity bias on tetrahedralized grids.
  }
\end{figure}

Barycenters are clearly continuous with respect to vertex positions and the cotagent Laplacian is negative semi-definite.
In particular, $\D_t^\text{bary} = \G_t^T \A^\text{bary}_t$ where $\A_t^\text{bary} = 
(\int_{\primal{t}} dV)
\, \I$ (see \reffig{ellipsoids}).
Barycenters always lie inside each respective simplex. This
immediately implies that the mass matrix is positive.
However, because tetrahedra volumes are distributed to each incident vertex
based on connectivity not geometry, the baricentric dual-volumes suffer strong
bias for a tetrahedralized regular grid (see \reffig{grid-3d}).

\subsection{Circumcentric dual-volumes}
\label{sec:circumcentric}
Circumcentric dual-volumes are defined by using the circumcenter of each simplex.
For interior vertices of a Delaunay mesh, this choice results in local dual
volumes matching the Voronoi diagram of vertices.
Circumcenters are a continuous function of vertex positions for non-denegerate meshes.

While not obvious \emph{a priori}, circumcentric dual-volumes result in a symmetric Laplacian matrix.
Unlike their 2D analogs for triangle meshes, circumcentric dual-volumes do not result, in general, in the same Laplacian matrix as the linear-FEM or barycentric dual-volume Laplacian:
\begin{equation}
  \L^\text{circum} \neq \L^\text{bary}.
\end{equation}
That is, in general, $\A^\text{circum}_t \neq \A^\text{bary}_t$ for a given tetrahedron $t$ as  $\A^\text{circum}_t$ is, in general, not diagonal. 
We could alternatively interpret a Laplacian built with circumcentric
dual-volumes as a piecewise-linear finite-element method applied to a mesh with
non-trivial piecewise-constant metric diffusion tensors $\A^\text{circum}_t \in
\R^{3 \times 3}$; in other words, $\A^\text{circum}_t$ determines a 
per-tetrahedron transformation $\J_t$ applied to each tetrahedron independently before applying
linear-FEM construction (see \reffig{ellipsoids}).
There is no guarantee that the per-tetrahedron transformations 
could be realized by a vertex deformation of the whole mesh in $\R^3$.

\begin{figure}[b!]
\includegraphics[width=\linewidth]{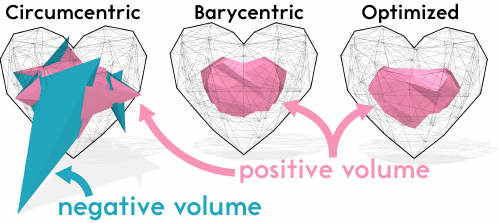}
  \caption{
    \label{fig:hearts}
    For non-Delaunay tetrahedral meshes such as this heart, circumcentric
    dual-volumes can result in negative entries in the mass matrix, due to self-intersecting,
    flipped-inside-out contributions from each element (left).
    Barycentric (center) and our optimized dual-volumes (right) are always positive.
  }
\end{figure}

For a regular grid, circumcentric dual-volumes are unbiased, coinciding with a finite-differences.
Unfortunately, if the mesh is non-Delaunay the mass-matrix is not guaranteed to
be positive.
Circumcenters may slip outside simplices, causing the boundary of the dual volume to self-intersect.
This may eventually cause the total signed volume of $\dual{i}
\cap t$ and even the total signed volume of $\dual{i}$ to be negative (see \reffig{hearts}).

\begin{figure*}
\includegraphics[width=\linewidth]{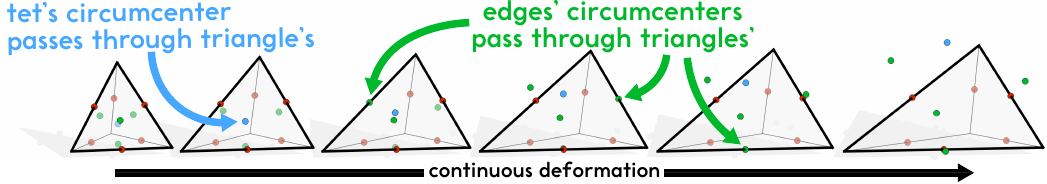}
  \caption{
    \label{fig:morph}
    \newhl{As a tetrahedron continuously deforms from equilateral to non-acute, the circumcenters of the tetrahedron 
    and its triangles pass through the triangle and edge circumcenters, respectively.
    This inspires our circumcentric snapping generalization of \cite{MeyerDSS02},
    which ensures continuity of dual-volumes but not symmetry of the Laplacian $\L$.
    }
  }
\end{figure*}

\subsection{\cite{AlexaHKS20} dual-volumes: barycentric snapping}
To combat negative volumes in 2D triangle meshes, Meyer et al.~\cite{MeyerDSS02}
proposed using circumcentric triangle centers \emph{except} when the triangle is obtuse.
In this case, Meyer et al. would snap the center to the midpoint of the edge opposite the obtuse angle.
Citing inspiration from the idea of center-filtering at a high-level, Alexa et
al.~\cite{AlexaHKS20} snap any centers lying outside its simplex to that
simplex's \emph{barycenter}.
By construction the mass-matrix is positive, but dual-volumes on a $\epsilon$-perturbed
regular grid are biased as every entry suddenly jumps in shape and size to the biased barycentric
dual-volume mass matrix (see \reffig{grid-continuity}).
Consider a simplex as it morphs from non-obtuse to obtuse: its circumcenter and
barycenter will not coincide during the transition.
Therefore, the dual volumes of \cite{AlexaHKS20} are not a continuous function
of vertex positions.

For \emph{triangle} meshes, the choice of triangle center does not affect the
definition of the Laplacian matrix \newhl{(by Stokes Theorem and that the
gradient is constant in each triangle. See, e.g., Sec.~3.3.4 in \cite{pmpbook})}.
As noted, edge midpoints are equivalent to barycenters, so the resulting
Laplacian matrix agrees with the \emph{symmetric negative semi-definite}
linear-FEM / barycentric dual-volume Laplacian.

For tetrahedral meshes, the choice of tetrahedron center also does not affect
the Laplacian \newhl{(again, by applying Stokes Theorem against a constant vector field in
each tetrahedron)} and of course edge circumcenters are also barycenters, but
obtuse triangle circumcenters snapped according to \cite{AlexaHKS20} \emph{will}
affect the Laplacian matrix.
%
%
As a result, the $\L^\text{A}$ will be non-symmetric in the presence of obtuse triangles.
Indeed, for a single tetrahderon with an obtuse triangle 
the resulting $\D_t^\text{A}$ does not satisfy $\D_t^\text{A} \G_t = \G_t^T (\D_t^\text{A})^T$.
We may solve for its analogous implied ``diffusion tensor'' \emph{post facto} as
\begin{align}
  \A^\text{A}_t = -\left(\G^T\right)^+ \D_t^\text{A} 
\end{align}
which in this case reveals that $\A^\text{A}_t$ is, in general, not symmetric and
$\A^\text{A} + (\A^\text{A})^T$ is not positive definite. This implies that 
we do not have, in general, a positive semi-definite Dirichlet energy formed
with $\L^\text{A}$ (see \reffig{fox}).
Across datasets of larger tetrahedral meshes, we experimentally find that $\L^\text{A}$ is
consistently significantly asymmetric. Often, $\L^\text{A}$ will have all
non-negative eigenvalues, but $\L^\text{A} + (\L^\text{A})^T$ is indefinite.
Meanwhile, other times $\L^\text{A}$ will even have non-trivially imaginary
eigenvalues, but $\L^\text{A} + (\L^\text{A})^T$ will be negative semi-definite.

%

\subsection{\cite{mullen2011hot} optimization over orthogonal dual volumes}
Orthogonal duals are a subset of simplex center construction where dual faces are geometrically orthogonal to primal edges.
The circumcentric dual-volumes of \refsec{circumcentric} are a special case of
orthogonal duals, but the space is much larger \cite{Glickenstein2005,Wardetzky07,
mullen2011hot}.
Orthogonal dual constructions for triangle meshes (and less often tetrahedral
meshes) have been considered in the past, primarily discussed through the lens
of discrete exterior calculus \cite{desbrun2005discrete}.

Mullen et al.~\cite{mullen2011hot} propose to optimize over the space of metrics
in each local neighborhood so that the dual is orthogonal by construction and
other properties such as mass positivity is encouraged by minimization.
Mullen et al.~may optimize both the metrics and the primal mesh simultaneously.
For our consideration, we will consider just the metric optimization.
Unfortunately, the optimization problem is non-linear and non-convex, requiring
a variant of gradient descent to find a local minimum.
This forgoes the guarantee of continuity with respect to vertex positions.
%
While they report \emph{reducing} the amount of negative volumes compared to
circumcentric duals, in their examples they do not eliminate negative volume entirely.
To our knowledge, there is no available open source implementation of this method
to provide explicit counter-examples here.


\subsection{Continuous dual-volumes via circumcentric snapping}
As a warm-up to our main result, we consider modifying the positive-volume
filtering of Alexa et al.~\cite{AlexaHKS20} to ensure continuity as tetrahedra morph from
acute to non-acute shapes \newhl{(we say a tetrahedron is ``acute'' if all dihedral angles are less than $\pi/2$ \cite{EppsteinSU04})}.

%
Unlike Alexa et al., the original center-filtering of Meyer et
al.~\cite{MeyerDSS02} ensures continuity with respect to vertex positions
because the triangle circumcenter is snapped to the \emph{circumcenter} of its subsimplex
(the midpoint of the edge).
We can generalize the choice of 
Meyer et al.~\cite{MeyerDSS02} to 3D tetrahedral meshes by 
snapping circumcenters of simplices recursively to subsimplices.
This choice ensures that dual-volumes are continuous with respect to vertex positions.
To see this, consider a non-obtuse simplex morphing into an obtuse one. At the moment that
its circumcenter exits the simplex it must coincide with a circumcenter of the
boundary subsimplex upon which it lies (because it is also equidistant to that
subsimplex's corners, see \reffig{morph}).

By design, snapping ensures that the mass matrix is positive, and snapping to circumcenters
ensures that the mass matrix of a regular grid remains unbiased like
circumcentric dual-volumes.
Unfortunately, circumcentric snapping --- like \cite{AlexaHKS20} --- does not guarantee symmetry or a positive
semi-definite Dirichlet energy.

\section{Optimized dual-volumes}
Drawing inspiration from Meyer et al. \cite{MeyerDSS02} and \cite{AlexaHKS20}, we propose to
utilize circumcentric dual-volumes when possible.
Unlike those methods, we will explicitly ensure a positive semi-definite Dirichlet energy \emph{by constraint}.

Vertex circumcenters are trivially well positioned, and edge circumcenters are
non-troublesome midpoints.
Meanwhile, the tetrahedra centers do not affect the
Laplacian matrix, so they can be snapped individually \emph{post facto}.
What remains are the triangle centers.

We propose to optimize the center location of each triangle to be as close as possible to its circumcenter, but to stay within their
respective triangles \emph{and} so that the implied diffusion tensor $\A_t$ of
each tetrahedron is positive definite.
To ensure the continuity of dual-volumes, two tetrahedra sharing a triangle
must agree on the triangle's center.
Thus we consider the following optimization over the centers of a tetrahedral
mesh with vertex positions in $\v_i \in \R^3$ for $i = 1, \ldots, n$ and
triangle indices gathered into $\F \in [1,n]^{k \times
3}$:
\begin{align}
  \label{equ:loss}
  \min_{\C \in \R^{k \times 3}}  & \quad \sum_{f} \left\|\o_f - \c_f\right\|^2  \\
  \label{equ:inside}
  \text{s.t.} & \quad \c_f \in \primal{f} &\forall\, f \\
  \label{equ:definite}
  \text{and}   & \quad (\G_t^T)^+ \D_t(\C) \preceq 0 \quad &\forall\, t 
\end{align}
where $\o_f$ is the (known) circumcenter of triangle $f$ and
$\c_f$ is the optimized center,
$\D_t(\C) : \R^{k \times 3} \rightarrow \R^{4 \times 3}$ is the deterministic and linear function which selects the centers of
the four triangles incident on tetrahedron $t$ from $\C$ and builds the
integrated divergence operator according to the construction in \refequ{discrete-div}.

Eqs.~\ref{equ:loss}-\ref{equ:definite} define a strictly convex semi-definite
programming (SDP) problem.
The problem always has a solution: triangle barycenters are feasible.
Therefore, solutions of this feasible strictly convex optimization are a
continuous function of its parameters (mesh vertex positions).
The constraint in \refequ{definite} ensures that the Laplacian matrix is
symmetric negative semi-definite, and thus the Dirichlet energy is positive
semi-definite.
The constraint in \refequ{inside} ensures that the mass matrix is positive.
\newhl{When a circumcentric dual-volume satisfies both these properties, our solution
will coincide, thus ensuring unbiased dual-volumes on regular grids.}
Perturbing the mesh vertices slightly will result in a continuous change,
degrading gracefully while maintaining definiteness and positivity (see
\reffig{grid-continuity}).

Large sparse semi-definite programs are tractable to solve, but rather slow.
Fortunately, we observe that simply constraining $\A_t$ to be \emph{symmetric} while
ensuring positive dual volumes (via \refequ{inside}) appears to always result in a positive semi-definite $\A_t$.
While we lack a \newhl{formal or symbolic proof, we constructed a numerical experiment on a unit tetrahedron to
hunt for an \emph{negative} definite $\A_t$ while keeping centers inside their
triangles: finding a result would amount to a counterexample to our claim.
However, the SDPT3 solver \cite{sdpt3} declared this counterexample-search
problem infeasible: agreeing with our hypothesis that symmetry and positive dual
volumes are sufficient conditions for positive definiteness.
Out of an abundance of caution, for a given mesh we can always verify whether the resulting symmetric $\A_t$ matrices are
positive semi-definite and, if not, resolve the full SDP using a large-sparse-SDP solver like Mosek \cite{mosek}.}
We have not encountered the need for this yet.

So, we propose to replace \refequ{definite} with the following linear equality constraints enforcing symmetry:
\begin{align}
  \label{equ:symmetric}
   \quad \D_t(\C)\, \G_t = \G_t^T\, \D_t(\C)^T \quad &\forall\, t.
\end{align}

\refequ{symmetric} appears to be a rank-6 constraint over the entries of $\D_t$.
However, after symbolical analysis, we find that this constraint is in fact only
rank 3 over the entries of $\C$.
Thus, each tetrahedron contributes 3 linear equality constraints (\refequ{symmetry-constraint}), touching only its four incident triangles' centers.

Combined with the quadratic loss in \refequ{loss} and the linear inequality constraints in
\refequ{inside}, we have a convex quadratic program (QP).
By re-parametrizing $\C$ as a linear function of barycenteric coordinates $\B \in \R^{k
\times 3}$, we can express the constraints in \refequ{inside} as
non-negative and partition of unity constraints on $\B$.
This large, sparse QP can be efficiently solved with a variety of standard
solvers \cite{mosek,nasoq,osqp,matlab}.

%

\subsection{Trivial extension to intrinsic dual-volumes}
While the value of intrinsic 2D triangulations has been made exceptionally
clear \cite{FisherSSB07,BobenkoS07,SharpSC19a,GillespieSC21a,LiuGCSJC23},
this value has perhaps not yet extended to 3D tetrahedral meshes.
Nevertheless the extension of our general centric construction --- and thus
also our optimized dual-volumes --- to intrinsic dual-volumes is a trivial
extension.
Our quantities and constraints are defined per-element and our optimization
variables can be expressed in terms of (intrinsic) barycentric coordinates.
To accommodate an intrinsic tetrahedralization --- where mesh edges have lengths
but no prescribed vertex positions --- one could simply map each tetrahedron to
a canonical tetrahedron with matching metric, compute relevant quantities and
constraints.
Each $\left\|\o_f - \c_f\right\|^2$ in \refequ{loss} would split into
contributions from each adjacent tetrahedron (expressed in barycentric coordinates).


\section{\newhl{Numerical Experiments}}

\begin{figure}
  \includegraphics[width=\linewidth]{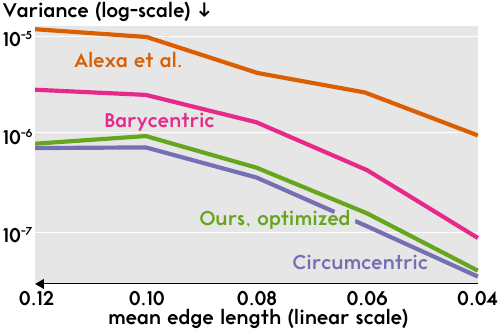}
  \caption{
    \label{fig:sphere-shells-variance-plot}
    \newhl{
      Compared to circumcentric, barycentric dual-volumes are known to produce more variance when used to
      solve a radially symmetric Laplace problem.
      The filtered centers of Alexa et al.~\cite{AlexaHKS20} result in even higher variance.
      Our optimized centers result in variance just above circumcentric duals.
      In all cases, variance decreases with mesh refinement.
    }}
\end{figure}

\newhl{
  In \reffig{sphere-shells-variance-plot},
  we replicate the ``Dirichlet energy minimization'' experiment from
  \cite{AlexaHKS20}, where we generate progressively refined tetrahedral meshes
  conforming to concentric spheres of radii 1.0, 0.75 and 0.5. We minimize
  $-\f^T (\L + \L^T) \f$ subject to boundary conditions of 1.0 and 0.0 on the
  inner and outer shells respectively, then measure variance of the vertex
  values on the middle shell.
  In our experiment, we generate sphere triangle meshes using centroidal Voronoi
  tesselation, then use Tetgen \cite{tetgen} to mesh in between them.
  As shown by Alexa et al., barycentric duals result in higher variance than
  circumcentric. However, the filtered centers proposed by Alexa et al.~result in
  even higher variance.
  Our optimized centers result in variance just above circumcentric duals.
  }

\begin{wrapfigure}[11]{r}{0.4\linewidth}
  \includegraphics[trim=0.5cm 0 0 0.0cm,width=\linewidth]{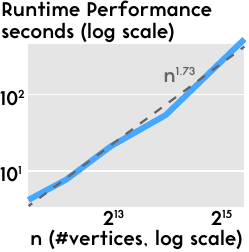}
\end{wrapfigure}
\newhl{
  We made almost no effort optimizing the runtime performance of our numerical implementation. We
  implemented the optimization of face centers in Matlab with lots of \texttt{for} loops. Nevertheless, the
  bottleneck is solving the large-sparse quadratic program (using Mosek \cite{mosek}).
  The number of variables in the optimization (three times the number of
  triangles $3k$) closely tracks $30\times$ the number of mesh vertices in the
  previous experiment.
  The inset plot shows that sub-quadratic performance in the number of
  vertices, far better than worst-case complexity for convex quadratic programs
  indicating that sparsity is paying off.
  Timings are recorded on a 2020 MacBook Pro laptop (2.3GHz Quad-core Intel
  Core i7, 32GB RAM).
}

\begin{figure}
  \includegraphics[width=\linewidth]{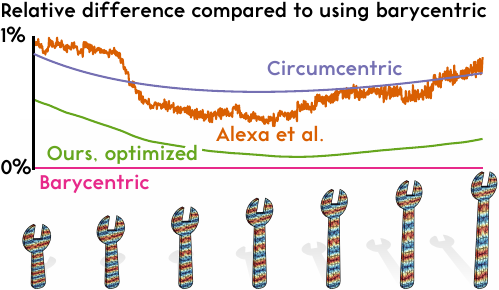}
  \caption{
    \label{fig:wrench}
    \newhl{
      Consider computing $\f^T \M \f$ as the vertex positions of a tetrahedral
      mesh change with its design (bottom)
      Barycentric, circumcentric, and our optimized dual-volumes are continuous
      function of mesh vertex positions, and so is this value computed with their respective mass matrices.
      The dual volumes of Alexa et al.~\cite{AlexaHKS20} are discontinuous, producing a noisy value as the design changes.
      The line plot (top) shows the relative difference of each method's value
      compared to that of barycentric dual volumes as the design changes.
    }}
\end{figure}

\newhl{
  Once constructed numerically, our Laplacian $\L$ and mass matrix $\M$ have the
  same sparsity and thus same runtime implications for downstreams tasks as
  others methods (but unlike alternatives with larger stencils, e.g., \cite{Bunge2021}).
  In \reffig{fox}, we compare the numerically computed eigenvectors of the Laplacian built from our dual volumes to that of Alexa et al.~\cite{AlexaHKS20}, whose
  lack of semi-definiteness manifests as numerical debris in the eigenvectors.
  Laplacian eigenmodes are a critical downstream task used in a variety of
  applications areas in geometry processing \cite{ValletL08,Rustamov07,SunOG09,OvsjanikovBSBG12,SharpACO22,BenchekrounZCGZJ23}.
}

\newhl{
  Our optimized dual-volumes are continuous with respect to vertex positions.
  In \reffig{wrench}, we consider a downstream design analysis task where the
  tetrahedral mesh connectivity of a wrench is transported along with procedural
  changes to its design \emph{\`a la} \cite{SchulzXZZGM17}.
  The squared norm of a sinusoidal function $\f$ is measured across the changing designs: $\f^T \M \f$.
  The line plots show relative differences of each method's value compared to
  that of barycentric dual volumes as the design changes.
  The dual volumes of Alexa et al.~\cite{AlexaHKS20} produce a noisy
  discontinuous function, while ours --- like barycentric and circumcentric ---
  are continuous with respect to change in vertex positions.
}

\section{Discussion \& Conclusion}
\label{sec:discussion}

%
\begin{wrapfigure}[10]{r}{0.4\linewidth}
  \includegraphics[trim=0.5cm 0 0 1.2cm,width=\linewidth]{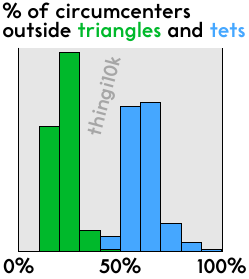}
\end{wrapfigure}
As casually observed by Alexa et al.~\cite{AlexaHKS20}, tetrahedral meshes are often not Delaunay in the wild.
\emph{Constrained} Delaunay meshers (e.g., \cite{tetgen,DiazziPVA23}) do not
enforce the local Delaunay criteria at constrained (e.g., boundary) faces.
Meanwhile, optimization-based meshers (e.g., \cite{tetwild,ftetwild}) will
trade Delaunay-ness for improved aspect ratios or other important mesh quality
measures.
Indeed, considering
the 10,000 outputs of \textsc{TetWild} run
on the Thingi10k dataset \cite{thingi10k}, we see that most tetrahedra
circumcenters lie outside, and roughly one quarter of face circumcenters lie
outside (inset).
Consequently, circumcentric dual-volumes result in
negative mass matrix entries for 74\% of the meshes across the dataset.
The barycentric snapping method of Alexa et al.~\cite{AlexaHKS20} results in
asymmetric Laplacian matrix $\L$ for 100\% of the models, and of these Cholesky
decomposition of $\L + \L^T$ fails on 57\%.

The general construction considered in this paper explicitly constructs the dual volume.
Besides making constraints and analysis explicitly clear, ensuring a geometric
realization could prove essential for applications which rely on sampling or
interpolation within dual-volumes.
Elcott et al.~\cite{ElcottTKSD07} use primal-dual meshes for fluids and generalized barycentric coordinates for interpolation.
Batty et al.~\cite{batty2010tetrahedral} extend this to non-conforming Delaunay
meshes, where circumcentric dual volumes are positive and convex shaped.
Our optimized dual-volumes are positive but not necessarily convex; on the other
hand, we do not require a Delaunay mesh, so meshes can be conforming.
While convenient for velocity interpolation, convexity is not strictly
necessary and generalized barycentric coordinates have since improved for non-convex
polyhedra \cite{JoshiMDGS07,HormannS08,ZhangDLPBHL14,ChangDH23}.

In this paper, we considered construction and optimization of dual-volumes
assuming a contract with the user that the primal mesh remains fixed.
Prior works have also considered (simultaneous) optimization of the primal mesh
or its intrinsic connectivity to improve Laplacian and mass-matrix operators
for 2D triangulations \cite{deGoes2014} and tetrahedral meshes
\cite{StroterMWF22} or both
\cite{mullen2011hot,AlexaHarmonic2019,AbdelkaderBEMMO20,AlexaConformingWeightedDelaunay2020,MitchellKMD23}.
Alexa~\cite{AlexaHarmonic2019} shows how unique the situation is for
triangulations in 2D and that many clear concepts or algorithms do not extend to 3D and
beyond.
While we believe our optimization generalizes in theory to higher dimensional
meshes, we leave this exploration to future work.

The space of orthogonal duals is too small to include the barycentric duals (and
thus linear FEM, the \emph{de facto} standard). 
Our general construction with simplex centers and space of optimization
explicitly includes these, while others do not \cite{Wardetzky07,mullen2011hot}.

When considering the orthogonal dual construction of the Laplacian for 2D
triangle meshes, the choice of triangle center at the circumcenter seems to have great
importance: Wardetzky et al.~\cite{Wardetzky07} write 
``the cotan weights ... arise from assigning dual vertices to circumcenters of
primal triangles'' \cite{Wardetzky07}).
However, in the more general view of constructing dual-volumes by connecting
simplex centers, the choice of triangle center is demonstrably irrelevant.
For the cotan weights to arise, only the choice of edge centers at midpoints
matters (see also \cite{pmpbook}).

In 3D, the situation is very different. Edge centers remain canonically
midpoints, but triangle centers are neither irrelevant nor canonically
determined.
Fortunately, they span an interesting enough space to find dual-volumes that
fulfil a variety of properties.
Like the triangle center in 2D, the tetrahedron center in 3D is irrelevant to
the Laplacian matrix, but  does affect the mass matrix.

The taxonomy of Wardetzky et al.~\cite{Wardetzky07} is not quite expressive
enough to capture the locality or lack thereof of our proposed approach.
The sparsity pattern of our Laplacian matrix matches the combinatorial mesh, but
the entries are effectively determined via a global optimization (similar to
\cite{mullen2011hot}).

Alexa et al.~\cite{AlexaHKS20} muse about linearly blending between barycentric and circumcentric
duals.
Using a single parameter for the entire mesh would result in a space that could ensure positivity, definiteness and
continuity, but not unbiased-ness on (local) grid structures.
It is not obvious how to spatially vary this blend so that properties are
ensured locally; our optimization works on a large space by considering
effectively two parameters per triangle (barycentric coordinates of its center).

Wardetzky et al.~\cite{Wardetzky07} propose a general construction of
symmetric 2D triangle mesh Laplacians by spanning a parameterization over the
metrics of each edge.
This is analogous to our observation that once $\nabla f$ is assumed to b e
piecewise constant in an element, what remains is the choice of divergence
operator $\D$ or diffusion tensor $\A$.
Perhaps the construction of Wardetzky et al.~could be generalized to tetrahedral
meshes, but it would seem to correspond to directly parameterizing $\A$ rather
than working with simplex centers.
By optimizing centers directly we ensure that dual-volumes are continuous across
triangles and embedded.
\newhl{In this paper, we focus on the effect of these properties on the
Laplacian and mass matrices. It would be interesting to explore our optimized
dual-volumes for other operators or uses (e.g., fluids \cite{batty2010tetrahedral}).
}

Employing convex-optimization to \emph{build} the Laplacian and mass matrices
will often be more expensive than a single subsequent sparse linear system
or truncated eigenvalue decomposition.
In some applications, ensuring the features in \reftab{feature} may be worth the
wait. 
In any case, our result also serves as an existence proof that we hope fuels
future research into more efficient methods.
Finally, any invocation of a feature chart like \reftab{feature} should be met
with a critical speculation about which other feature columns should be
considered in the future.
We hope the analysis and novel method presented here will continue to fuel
research into dual-volumes.
\newhl{To this end, we have open-sourced implementation as part of the \final{gptoolbox \cite{gptoolbox}} library.}

\section*{Acknowledgements}
\newhl{
Our research is funded in part by NSERC Discovery (RGPIN–2022–04680), the
Ontario Early Research Award program, the Canada Research Chairs Program, a
Sloan Research Fellowship, the DSI Catalyst Grant program and gifts by Adobe
Inc.
We are grateful for the feedback of the Banh Mi weekly meeting attendees at
University of Toronto, insightful conversations with Oded Stein \& Nick Sharp,
and early draft comments from Towaki Takikawa, Abhishek Madan, \& Eitan
Grinspun.
The ``Fennec Fox'' model in \reffig{fox} is by Physics\_Dude by and used
according to the CC BY 4.0 license.
}

\printbibliography

\end{document}